
\input harvmac

\nref\rBethe{H. A. Bethe, Z. Phys. 71 (1931), 205.}
\nref\rBRone{V.~V. Bazhanov and N.~Yu. Reshetikhin,
 Int. J. Mod. Phys. A4 (1989) 115.}
\nref\rBRtwo{V.~V. Bazhanov and N.~Yu. Reshetikhin, J.~Phys.~A23 (1990), 1477.}
\nref\rBRthree{V.~V. Bazhanov and N.~Yu. Reshetikhin, Prog. Theor. Phys. Suppl.
102 (1990) 301.}
\nref\rKuniba{A. Kuniba, Nucl. Phys. B389 (1993) 209.}
\nref\rABF{G. E. Andrews, R. J. Baxter and P. J. Forrester, J. Stat. Phys. 35
(1984) 193.}
\nref\rJKMO{M. Jimbo, A. Kuniba, T. Miwa and M. Okado, Comm. Math. Phys., 119
(1988), 543.}
\nref\rADMone{G. Albertini, S. Dasmahapatra and B.~M. McCoy, Int. J.
 Mod. Phys. A7, Suppl. 1A (1992) 1.}
\nref\rADMtwo{G. Albertini, S. Dasmahapatra and B.~M. McCoy, Phys. Lett. A
 170 (1992) 397.}
\nref\rFab{F. H. L. Essler, V. E. Korepin and K. Schoutens, J. Phys. A 25
(1992), 4115.}
\nref\ryangs{C. N. Yang and C. P. Yang, J. Math. Phys. 10 (1969), 1115.}
\nref\rKedMc{R. Kedem and B.~M. McCoy, J. Stat. Phys. (in press).}
\nref\rDKMM{S. Dasmahapatra, R. Kedem, B.~M. McCoy and E. Melzer
 preprint, hep-th.}
\nref\rKKMMone{R.Kedem, T.~R. Klassen, B.~M. McCoy, and E. Melzer,
 Phys. Lett. B 304, 263.}
\nref\rKKMMtwo{R.Kedem, T.~R. Klassen, B.~M. McCoy, and E. Melzer,
 Phys. Lett. B (submitted).}
\nref\rDKKMM{S. Dasmahapatra, T. Klassen, R. Kedem, B.~M. McCoy and E. Melzer
 preprint, hep-th.}
\nref\rNRT{W. Nahm, A. Recknagel, M, Terhoeven, Bonn preprint, hep-th.}
\nref\rTer{M. Terhoeven, Bonn preprint, hep-th.}
\nref\rKNS{A. Kuniba, T. Nakanishi, J. Suzuki, Harvard preprint HUPT-92/A069.}
\nref\rRS{B. Richmond and G. Szekeres, J. Austral. Soc. (Series A) 31
(1981), 362.}
\nref\rBGS{A. Berkovich, C. Gomez and G. Sierra, preprint, hep-th.}
\nref\rFatZam{V. A. Fateev and A. B. Zamolodchikov, Phys. Lett. A 92 (1982),
37.}
\nref\rAl{G. Albertini, in preparation.}
\nref\rTak{M.~Takahashi, Prog. Theo. Phy. 46 (1971), 401.}
\nref\rstd{M. Takahashi and M. Suzuki, 48 (1972), 2187, M. Gaudin, Phys. Rev.
Lett., 26 (1971), 1301.}
\nref\rJimbo{M. Jimbo,  Lett. Math. Phys. 10,(1985) 63.}
\nref\rVerlinde{Erik Verlinde, Nucl. Phys. B 300 (1988), 360.}
\nref\rAGS{L. Alvarez-Gaum\'e, C. Gomez and G. Sierra, Nucl. Phys. B 330
(1990), 347.}
\nref\rGKO{P. Goddard, A. Kent and D. Olive, Comm. Math. Phys., 103 (1986),
105.}
\nref\rDdV{C. Destri and H.~J. de Vega, Nucl. Phys. B 385 (1992), 361.}
\nref\rKirdilog{A.~N. Kirillov, Dilogarithms and spectra in conformal field
theory, preprint, hep-th.}
\nref\rKirillov{A.~N. Kirillov, Fusion rules and Verlinde's formula,
preprint, hep-th.}
\nref\rKircount{A.~N. Kirillov, J. Sov. Math., 36 (1987).}
\nref\rGN{F.~M. Goodman and T. Nakanishi, Phys. Lett. B262 (1991), 259.}
\nref\rComb{S. Kerov, A.~N. Kirillov and N.~Yu. Reshetikhin, J. Sov. Math.
41 (1988), 916, A.~N. Kirillov and N.~Yu. Reshetikhin, J. Sov. Math. 41
(1988), 925, A.~N. Kirillov and N.~Yu. Reshetikhin, Lett. Math. Phys. 12,
(1986), 500.}
\nref\rMac{I. G. Macdonald, {\it Symmetric Functions and Hall Polynomials},
Clarendon Press, Oxford, 1979.}
\nref\rPS{V. Pasquier and H. Saleur, Nucl. Phys. B 330 (1990), 523.}
\nref\rGanchev{P. Furlan, A. Ch. Ganchev and V. B. Petkova, Nucl. Phys. B
(1990), 205.}
\nref\rLusztig{G. Lusztig, Geom. Ded. 35 (1990), 89.}
\nref\rLP{J. Lepowsky and M. Primc, {\it Structure of standard modules for the
affine Lie algebra} $A_1^{(1)}$, Cont. Math., Vol. 46 (AMS, Providence, 1985).}
\nref\rKNlevrank{A. Kuniba and T. Nakanishi, {\it in} Modern quantum field
theory, ed. S. Das, A. Dhar, S. Mukhi, A. Raina and A. Sen (World Sci.,
Singapore, 1991). }
\nref\rSA{H. Saleur and D. Altsch\"uler, Nucl. Phys. B 354 (1991), 579.}
\nref\rGepner{D. Gepner, Comm. Math. Phys., 141 (1991), 381.}
\nref\rcryone{M. Kashiwara, Comm. Math. Phys., 133 (1990), 249.}
\nref\rcrytwo{M. Jimbo, K. Misra, T. Miwa and M. Okado, Comm. Math. Phy. 136,
(1991), 543.}

\Title{\vbox{\baselineskip12pt\hbox{ICTP Preprint}
\hbox{hep-th/yymmnn}}}
{\vbox{\centerline{String Hypothesis and Characters of Coset CFTs}}}

\vskip 10mm
\centerline{ \rm Srinandan Dasmahapatra}
\vskip 10mm
\centerline{\it International Centre for Theoretical Physics}
\centerline{\it P.~O. Box 586, Strada Costiera 11}
\centerline{\it I-34100 Trieste, Italy.}

\vskip 20mm
\centerline{\bf Abstract}
\vskip 5mm

We present an algorithm for the construction of the branching functions in
the vacuum sector for affine Lie algebras based on the string hypothesis
solution to a system of Bethe equations for generalized RSOS models.
We also mention how the ground state structure and features of the excitation
spectra like the Brillouin zone schemes of these models (and those in the
same universality classes) can be extracted from combinatoric arguments and
encoded in Lie algebraic terms.

\Date{\hfill 4/93}
\vfill\eject

The exact solution of an exactly solvable model in 1+1 or 2+0 dimensions
usually implies the parametrization of the dynamical quantities of interest
(like the energy and momentum eigenvalues, and often, even the eigenvectors)
in terms of the solutions of a coupled set of transcendental equations, called
Bethe equations \rBethe. In the critical case, they are  usually written in
terms of trigonometric functions, and for the generalized RSOS models, they
are of the form
\eqn\bethe{\Biggl[{\sinh\bigl({\pi\over2 L}(\lambda_j^{(a)}+i(s \omega_p |
\alpha_a ))
\bigr)\over\sinh\bigl({\pi\over2 L}(\lambda_j^{(a)}-i(s \omega_p | \alpha_a ))
\bigr)}\Biggr]^N=\Omega_j^{(a)}\prod_{b=1}^r \prod_{k=1}^{N_b}
{\sinh\bigl({\pi\over2 L}(\lambda_j^{(a)}-\lambda_k^{(b)}+i (\alpha_a |
\alpha_b))
\bigr)\over\sinh\bigl({\pi\over2 L}(\lambda_j^{(a)}-\lambda_k^{(a)}-i
(\alpha_a | \alpha_b))\bigr)}.}
\eqn\tot{N_a=N s \bigl[C^{-1}\bigr]_{a p},}
where $N$ is the size of the lattice $\alpha_a$ are the simple roots,
$\omega_p$ are the fundamental weights, $(\cdot | \cdot)$ is the canonical
bilinear form on the dual
to the Cartan subalgebra, $C$ is the Cartan matrix, $L=l+g$, for integer
$l$, $g$ is the dual
Coxeter number and $s$ characterizes the type of fusion.
It should be mentioned that these have been derived in \rBRtwo  only for the
JKMO \rJKMO  models that generalize the ABF \rABF  models to the $A_n^{(1)}$
case (and their fusion hierarchies), and have
only been postulated for the other cases. (See \rKuniba  for a discussion.)
Here $\Omega^{(a)}$ is some phase factor that influences what the allowed
solutions are. Since all the quantities of physical interest are parametrized
by solutions to these equations (which are very difficult to solve), the
standard practice is to assume a particular form for the roots called the
``string hypothesis'' \rBethe and hope that the equations that are set up
in the thermodynamic limit to compute the spectrum or the free energy are
consistent with this ansatz. In fact, one does more -- one makes the tacit
assumption that important physical quantities are ``diagonalized in the
string basis,'' and therefore classifying the spectrum in terms of the string
form of the roots of the Bethe equations becomes significant.
In what follows, the string hypothesis (as in \rBRone \rBRtwo and \rKuniba)
can be written as:

\eqn\str{\lambda_j^{(a)}=\lambda + i t_a^{-1}  (j+1-2j_1), ~~~1\leq j_1\leq j}
with $t_a^{-1}=(\alpha_a|\alpha_a)/2$. $\lambda$ is real and denotes the
center of the string and $j$
denotes its length, which is further assumed to satisfy $1\leq j\leq t_a l$.
We shall therefore impose
$\sum_{j=1}^{t_a l} j M_j^{(a)}=N_a$, where $M_j^{(a)}$ denotes the number
of strings of colour (a), length $j$ and $N_a$ is given by \tot.
These assumptions were made in the bulk in \rBRone  (and subsequently in
\rBRtwo  and  \rKuniba), but here we shall impose these on finite-size
lattices. (Of course, the same thermodynamic limit can be recovered.)
Despite the well-known examples of non-string like behaviour
(see \rADMone, \rFab for a list of references) the results presented here
do seem rather striking, and it would be interesting to understand to
what extent these results are related to the actual models one would
like to solve.

As always, we multiply out the Bethe equations for the components of each
string, and end up with equations for the real parts of the roots.
We then take the logarithm of \bethe  so that the integer branches are
distinct.
For the simply-laced algebras, the formulas look simpler, (with all $t_a=1$)
and for this case, we have
\eqn\log{N t_{j, s}^{(a)}(\lambda_{\mu}^{j (a)})~=~2 \pi i I_{j \mu}^{(a)} +
\sum_{b=1}^r \sum_{k=1}^{l} \sum_{\nu=1}^{M_k^{(b)}} \Theta_{j k}^{(a b)}
 (\lambda_{\mu}^{j (a)}-\lambda_{\nu}^{k (b)}),}
where $\lambda_{\mu}^{j (a)}$ labels the center of the $\mu$-th string of
length $j$ and color $(a)$ which is a root of eq. \bethe. The functions
$t_{j, s}^{(a)}$ and
$\Theta_{j k}^{(a b)}$ are defined below.
\eqn\tjs{t_{j, s}^{(a)} (\lambda)=\delta_{a p}\sum_{k=1}^{min(j,s)}
f(\lambda;|j-s|+2k-1),}
\eqn\theta{\eqalign{\Theta_{j k}^{(a b)} (\lambda)&=\delta_{a b} \Biggl(
f(\lambda;|j-k|)+2\sum_{i=1}^{min(j,k)-1} f(\lambda;|j-k|+2i) +
f(\lambda;j+k)\Biggr)
\cr&~~~~~~~-I_{a b}\Biggl(\sum_{i=1}^{min(j,k)} f(\lambda;|j-k|+2i-1)\Biggr),}}
\eqn\f{f(\lambda;n)=~{1\over2\pi i}\ln\Bigl({\sinh {1\over2}{\pi\over L}(i n
-\lambda)\over \sinh {1\over2}{\pi\over L} (i n +\lambda)}\Bigr), }
for integer values of $n/L$, and is $0$ otherwise. $I_{a b}$ is the incidence
matrix of the respective Dynkin diagrams.
(For the non-simply laced cases, the branches are chosen following the same
principle \rADMone, but they look a little more complicated, and for the
case of $B_n$ and $F_4$, an extra prescription is required.)
We then follow the (standard) procedure \rstd  of defining
\eqn\zfun{Z_j^{(a)}(\lambda)\equiv t_{j, s}^{(a)}(\lambda) - {1\over N}
\sum_{b=1}^r \sum_{k=1}^{t_b l-1} \sum_{\nu=1}^{M_k^{(b)}} \Theta_{j k}^{(a b)}
(\lambda, \lambda_{\nu}^{k (b)}),}
so that the (half-) integers $I_{j \mu}^{(a)}$ satisfy
\eqn\zrts{Z_j^{(a)}(\lambda_{\mu}^{j (a)})={I_{j \mu}^{(a)}\over N}.}
If we assume that $Z_j^{(a)}(\lambda)$ is monotonic, then the range of the
integers is set by taking the difference of the limiting values, $Z_j^{(a)}
(\pm\infty)$. For a given choice of the phase factor $\Omega_j^{(a)}
(=e^{(2\pi s i/L)}$ where the ground state lies),
the first striking consequence of these assumptions is that
for strings of length $l$, the range of integers $\Delta I_l^{(a)}$
coincides with the total number of strings of that length, $M_l^{(a)}$,
corresponding to ``no holes'' as seen in the bulk calculations of \rBRone
and \rBRtwo. This reinforces our understanding that all the statements
made in the bulk using densities can be upgraded to finite lattices. This
will be exploited further later. Since the $l$-strings do not contribute to
the counting of states, we can eliminate $M_l^{(a)}$ using the sum rule
\tot. The range of integers associated with the other strings are:
\eqn\rangem{\Delta I_j^{(a)}= N \delta_{a p} \bigl[ C_{A_{l-1}}^{-1}
\bigr]_{j s}
+ M_j^{(a)} - \sum_{b=1}^r \sum_{k=1}^{l-1} (C_{\cal G})_{a b}
\bigl[ C_{A_{l-1}}^{-1}\bigr]_{j k} M_k^{(b)},\hbox{\hskip.5cm}j<l,}
where $C_{\cal G}$ is the Cartan matrix of the (simply-laced) Lie algebra
$\cal G$ and $C_{A_{l-1}}^{-1}$ is the inverse Cartan matrix of $A_{l-1}$.
This gives a combinatorial count of the number of states of the form
described above,
\eqn\count{ \sum_{\{M_j^{(a)}\}} \prod_{j,a} {\Delta I_j^{(a)}\atopwithdelims
() M_j^{(a)}}.}
We shall use the equality of binomial coefficients
\eqn\bin{ {A\atopwithdelims () B}={A\atopwithdelims () A-B}}
to define a new variable
\eqn\defn{ N_j^{(a)}=\Delta I_j^{(a)} - M_j^{(a)}, }
which counts the number of holes. We shall have some more to say about
this shortly.

To make the idea of upgrading bulk quantities to finite lattices concrete,
we define densities for strings and holes \ryangs  by requiring their
integrals over
all $\lambda$ to be exactly equal to the number of strings and holes divided
by $N$, the size of the system. This is the precise statement that rules the
interpolation between the bulk and finite lattices. In the notation of
\rKuniba,
\eqn\densities{ \hat{\rho}_j^{(a)}(0)=M_j^{(a)}/N , ~~~ \hat{\sigma}_j^{(a)}(0)
=N_j^{(a)}/N,}
where $\hat{f}(0)$ denotes the Fourier transform of $f$ for zero argument.
In the thermodynamic limit, because solving the integral equations involves
taking derivatives on the densities, one can choose the branches of the
logarithms (of the Bethe equations) freely. What we see is that there exists
a choice of branch on the finite lattice that makes the translation of
statements true in the bulk to finite lattices possible.

Instead of setting up the equations for the thermodynamics here, it is
more convenient to directly quote the result from ref. \rKuniba. We perform
the trivial steps of setting $x$ (the Fourier transform variable) to zero,
and applying the definition \densities to eqn. (2.16) of ref. \rKuniba to get
\eqn\generaln{N_j^{(a)}=N \delta_{a p} [C_{A_{t_p l-1}}^{-1}]_{j s} -
\sum_{b=1}^r \sum_{k=1}^{t_b l-1} K_{a b}^{j k} M_k^{(b)},}
where
\eqn\defK{K_{a b}^{j k} \equiv (\alpha_a | \alpha_b) \{ \hbox{\rm min}(t_b j,
t_a k) - {jk\over l}\}.}
We can also solve for $M_j^{(a)}$ in terms of $N_j^{(a)}$ to get
\eqn\generalm{M_j^{(a)}=N\sum_{k=1}^{t_p l-1} (K^{-1})_{a p}^{j k}
[C_{A_{t_p l-1}}^{-1}]_{k s} - \sum_{b=1}^r \sum_{k=1}^{t_b l-1}
(K^{-1})_{a b}^{j k} N_k^{(b)},}
where $K^{-1}$ is defined by
\eqn\defKin{\sum_{c=1}^r \sum_{m=1}^{t_c l-1} (K^{-1})_{a c}^{j m}
K_{c b}^{m k}=\delta_{a b} \delta_{j k}.}
In the above, the index $t_a l$ never shows up because there are never any
holes in this sector, and the sum rule \tot  is used to eliminate the
dependence on $M_{t_a l}^{(a)}$. Note that for the simply-laced cases, $K$
factorizes into a ``level'' and a ``rank'' piece, and \generaln  reduces to
\rangem  using \defn.

We can thus upgrade \count  to the general case of all untwisted affine Lie
algebras, where the number of states is of the form
\eqn\gencount{{M_j^{(a)}+N_j^{(a)}\atopwithdelims () M_j^{(a)}}=
{M_j^{(a)}+N_j^{(a)}\atopwithdelims () N_j^{(a)}}.}
Once the counting problem is set up, it is possible to extract information
about the ground and excited states by relying on the (tacit) assumption
that the physical quantities of interest are ``diagonalized in the string
basis,'' without even specifying the dependence of the energy of the
system on the roots of the Bethe equations! If the one-dimensional
hamiltonians are derived from the transfer matrix of a 2-dimensional
classical statistical mechanical model with positive Boltzmann weights,
then the Perron-Frobenius theorem forces the ground state to be unique. For
the RSOS models, we are fortunate in having a situation where both the largest
and smallest eigenvalue of the 1-dimensional hamiltonian correspond to
regimes in the 2-dimensional model where the Boltzmann weights are
positive. Therefore the ground states for the hamiltonians with either a
positive or a negative overall sign are unique.

One candidate is, of course, the sea of $t_a l$ strings, because it has no
holes. For the other candidate, we have to solve for the equation
$N_j^{(a)}=0$ which implies that the number of strings in that state is
\eqn\ground{M_j^{(a)}=N\sum_{k=1}^{t_p l-1}
(K^{-1})_{a p}^{j k} [C_{A_{t_p l-1}}^{-1}]_{k s} .}

Since the ground state must, by definition, be stable under an arbitrarily
large but finite number of excitations, we can see that the natural
variables to discuss the excitations are the $M$s for the ground state of
$t_a l$ strings, and the $N$s for the case where the ground state is
charecterized by the solution to \ground.
The momentum eigenvalue is written as a sum of the left hand sides of
\log \rBRone \rBRtwo \rKuniba (as in Bethe's solution of the isotropic
spin $1/2$ spin chain \rBethe). Thus, the
sum of the integers for any set of roots gives the total momentum (with the
appropriate $2\pi\over N$ taken into account) of the state
that these roots correspond to. Therefore we see that for every term in
$\Delta I_j^{(a)}=M_j^{(a)}+N_j^{(a)}$ for which the term proportional to
$N$ does not have support, the contribution of these strings (or holes) to
the momentum of the eigenstate at order one ($N^0$) must necessarily vanish.
For a system with no mass gap the energy contribution must consequently be
zero to order one. However, they could (and do) contribute to the spectrum
at order $1/N$. These have been termed ``ghost'' excitations in \rDKMM.
The strings (and holes) whose ranges are macroscopic ({\it i.e.} have a term
proportional to $N$) constitute the order one excitation spectrum, and
the coefficient of $N$ encodes the Brillouin zone scheme of these excitations.

To make these discussions more concrete, let us specialize to the
simply-laced case (all $t_a=1$) where $K_{a b}^{j k}=(C_{\cal G})_{a b}
[C^{-1}_{A_{l-1}}]_{j k}$.Recall that for one of the two
regimes ($\epsilon=+1$) in \rBRone and \rBRtwo, the state filled with $l$
strings is the ground state, and all the other strings which have
a macroscopic range (i.e., for which the Kr\"onecker delta in the term on the
r.h.s. of \rangem proportional to $N$ has support) have non-zero energy.
As an example, the 3-state Potts model is in the same universality class
as (and in fact can be obtained by orbifolding) the RSOS model
corresponding to $SU(2)$, with $l=4$. We can read off the Brillouin zone
scheme of \rADMtwo  from eq. \rangem. (Technical aside: also of interest is
the direct
correspondence between the ``zeroes of Q'' as studied here, and the ``zeroes
of T'' as studied in \rADMone, \rADMtwo. For $l=4$, 2$M_1$ corresponds to the
number of $2$-strings, 2$M_2$ to the number of $ns$, each dressed by $2$ plus
roots, 2$M_3$ to the number of $-2$-strings each dressed by $4$ plus roots and
2$M_4$ to the (even) number of minus roots each dressed by $3$ pluses, in the
terminology of \rADMone. The counting works out analogously, as does the
correspondence with the ground state structure and excitation spectrum.
It would thus be of interest to study this in more detail.)

For the other choice of ground state, which corresponds to the case
$\epsilon=-1$ in \rBRtwo, we find, from eqn. \ground,  that
the state with only $M_s^{(a)}$ non-zero has no
holes, and is the vacuum of the theory, and these are the only type of strings
that contribute to the energy and momentum at order 1 (in $N$). In the
relevant hole variables, the integer ranges look like
\eqn\rangen{\Delta I_j^{(a)}= N \delta_{s j} \bigl[ C_{\cal G}^{-1}
\bigr]_{a p}
+ N_j^{(a)} - \sum_{b=1}^r \sum_{k=1}^{l-1} (C_{\cal G}^{-1})_{a b}
\bigl[ C_{A_{l-1}}\bigr]_{j k} N_k^{(b)},
}
and makes the last remark transparent.

At this point we make a note that following the correspondence of \rGN,
if the string hypothesis gives an exhaustive count of all the states in
this (ground state) sector,
these sums of products of binomial coefficients
are precisely the multiplicities of the conformal blocks
made up of tensoring $N$ copies of $s \omega_p$ representations and projecting
on the singlet, and can be computed using Verlinde's formula \rVerlinde.
The inhomogeneous chain will give the analogous count for tensoring arbitrary
representations. This has also been observed in \rKirillov for $su(2)$. In the
simplest case of the
fundamental representations of $su(2)$, these numbers are indeed the number of
paths leading to the singlet on a truncated Bratelli diagram as in eq. (1.21)
of \rPS (see also \rGanchev). The binomial sums can be performed
in terms of generating functions, and the answer for $su(2)$ (following
the Appendix of  \rTak) is expressed in terms of
contour integrals over Chebyshev polynomials, suggesting a possible
connection with the fusion rings of \rGepner. Not surprisingly, the pole
terms that contribute are precisely those algebraic numbers that satisfy the
infinite spectral parameter thermodynamic Bethe ansatz equations that
determine the arguments of the
dilogarithms that give the central charge \rKirdilog \rKuniba \rNRT.
For the higher rank algebras, we expect to encounter the group characters
(following \rKircount).  However, this
analysis is still incomplete and we hope to report on this in the near
future, where we shall also present some details on Yangian characters.
(It is easy to see that the formulas in the appendix of \rKuniba can
be obtained from our combinatoric formulas by setting $l$ to infinity,
which is the ``classical'' limit of the ``quantum group.'' With these
explicit formulas the truncation condition in \rKuniba can presumably
be verified.)

We are now in a position to construct the branching functions for affine
Lie algebras. It is seen in numerical studies (which provided the background
for ref. \rADMone), that within each class of states with a paricular root
content, the sum of the absolute values of the integers give a good estimate
of the (approximate) degeneracy of the levels, i.e., those states with the
same value for the sum of the absolute value of the integers had energy
eigenvalues that were almost equal. This is reminiscent of the conformal
field theory definitions of energy and momentum being the sum and
differences of the eigenvalues of $L_0$ and $\bar L_0$.
One could then presume that this
degeneracy would become exact in the thermodynamic limit, and counting
degenerate states could be achieved by keeping track of these integers.
There are important variations on this
observation, however, in that the set of strings that do not have macroscopic
ranges actually contribute both a positive and
a negative amount in the energy eigenvalue at order $1/N$.
It is because these integers (actually, the momenta that correspond to
them) characterize the ``fermionic'' (distinct integers!)
quasiparticle spectrum, the resulting modular forms constructed by
$q$-counting or, alternatively, building up the partition function in a
particular sector, are called ``fermionic'' in \rKedMc, \rDKMM, \rDKKMM.

We shall distribute these integers with powers of $q$ by using the
the symmetric (under $q \rightarrow q^{-1}$) $q$-numbers (as in \rJimbo)
\eqn\qnum{\eqalign{ \{n\}&={(q^{n/2} - q^{-n/2})\over(q^{1/2} - q^{-1/2})},
{}~~~~~n \in Z,\cr
\{n\}!&=\prod_{i=1}^n \{i\},\cr
{A \atopwithdelims \{ \}  B}_q&={\prod_{i=1}^B \{A-B+i\} \over \{B\}!},}}
and construct the objects
\eqn\charsb{\eqalign{\chi_p&=\sum_{\{M_j^{(a)}\}} \prod_{j,a}
{\Delta I_j^{(a)}\atopwithdelims \{ \} M_j^{(a)}},\cr
\chi_h&=\sum_{\{M_j^{(a)}\}} \prod_{j,a}
{\Delta I_j^{(a)}\atopwithdelims \{ \} N_j^{(a)}}.}}

\par In order to focus on either chiral half of the spectrum, it is more
instructive to write these objects in terms of $q$-binomial coefficients,
which are polynomials in $q$ and are defined as
\eqn\qbin{ {A\atopwithdelims [] B} ~= ~~{(q;q)_A \over (q;q)_{A-B}
(q;q)_B},}
where
\eqn\qfac{(q;q)_A~=~\prod_{j=1}^A (1-q^j),}
(non-zero only for integers, $A$ and
$B$, with $0\leq A\leq B$). These are then appropriately symmetrized under
$q\rightarrow q^{-1}$ by dividing by one-half of the degree of the polynomial.
We thus have
\eqn\charsa{\eqalign{\chi_p&=\sum_{\{M_j^{(a)}\}} \prod_{j,a}
q^{-{1\over2} M_j^{(a)} N_j^{(a)}}
{M_j^{(a)} + N_j^{(a)}\atopwithdelims [] M_j^{(a)}},\cr
\chi_h&=\sum_{\{M_j^{(a)}\}} \prod_{j,a}
q^{-{1\over2} M_j^{(a)} N_j^{(a)}}
{M_j^{(a)} + N_j^{(a)}\atopwithdelims [] N_j^{(a)}},
}}
where we write $\chi_p$ only in the $M_j^{(a)}$ variables and $\chi_h$
only in the $N_j^{(a)}$ variables, using \generaln and \generalm.
Since large values of $\lambda$ correspond to small momentum contributions
and therfore, low energy states, and the integers have a monotonic
dependence on $\lambda$, the largest integers correspond to the lowest energy,
and we assign zero momentum to the (half-)integer $N/2$ that lies in the
power of $q$ multiplying the $q$-binomial coefficients.

Now the $\chi_p$ and $\chi_h$ look like
\eqn\charfin{\eqalign{\chi_p&=\sum_{\{M_j^{(a)}\}}
q^{-{1\over2} M\cdot K\cdot M} \prod_{j,a}
{{N\over2}\delta_{a p} K_{a p}^{j s} +  M_j^{(a)} + \sum_{b=1}^r
\sum_{k=1}^{t_b l -1} K_{a b}^{j k} M_k^{(b)}\atopwithdelims [] M_j^{(a)}},\cr
\chi_h&=\sum_{\{N_j^{(a)}\}}
q^{-{1\over2} N \cdot (K^{-1})\cdot N} \times \cr & \times\prod_{j,a}
{N\sum_{k=1}^{t_p l-1} (K^{-1})_{a p}^{j k}
[C_{A_{t_p l-1}}^{-1}]_{k s} - \sum_{b=1}^r \sum_{k=1}^{t_b l-1}
(K^{-1})_{a p}^{j k} N_k^{(b)} + N_j^{(a)}\atopwithdelims [] N_j^{(a)}},
}}
with the condition that
\eqn\limq{\lim_{N\rightarrow\infty} {N\atopwithdelims [] m}~=~{1\over (q;q)_m}
.}

These are the branching functions (in the vacuum sector) corresponding to the
conformal field theories constructed as cosets ({\it a la} GKO) \rGKO.  The
list of branching functions these $q$-series expansions correspond to are
given in \rKuniba. The central charges which can be obtained by taking
the $q\rightarrow 1$ limit as in \rRS, \rNRT  and \rKKMMtwo are also listed
there. One can read off a large
number of non-trivial equivalences between models based on different groups
and at various levels of fusion and roots of unity (level-rank dualities
\rKNlevrank \rSA).
To come back the 3-state Potts again, for example, we see that the
$G_2^{(1)}$ level 1 model for $p=2$ (arbitrary $s$) gives the Lepowsky-Primc
\rLP, \rKKMMone, \rTer form for the $c=4/5$ character. One could surely find
lots of other interesting examples.

It is amusing that the
$q\rightarrow 1$ limit for the symmetric $q$-binomial (the $ \{ \} $form) gives
the total count of the states, while by pushing the focus onto the low-lying
states (the [ ] form), one can extract the entropy of the low lying states
in the same limit.

In \rDdV, there is a simple prescription for restricting to the type II
\rPS  \rLusztig  integrable representations which are relevant for these
models -- namely,
the roots of the Bethe equations are non-singular. They also propose that
in order to go to a different sector one has to send a string to infinity.
The only change in the above formulas (in the $M$ language) would be to
shift the $M_j^{(a)}$ variable by $\delta_{j i}$, where $i$ refers to the
particular length of the string that is sent off to infinity. This
immediately gives the Lepowsky-Primc form \rLP \rKKMMone \rTer for {\it all}
the
$A_1^{(1)}$ branching functions (after setting ${\cal G}$ to su(2)).
The restriction on the summation variables reflect the total number of
strings in each sector. We conjecture that the same procedure for the higher
rank cases would give all the branching functions of the generalized
parafermionic theories. The transformation of this operation to the hole
language would give the branching functions in the other regime, but the
exact principle behind the restrictions on the integers seems unclear so
far.

It is intriguing that in ref \rKNS, the authors
produced their character formulas by considering the branches of the
dilogarithm, whereas we seem to do so with the ordinary logarithm(ic form
for the Bethe equations). Further
relations between these two approaches might lead to a greater
understanding regarding the solutions of the Bethe equations.

For the elliptic case considered in \rBRthree, the essential steps
of setting up the $q$-series representation can easily be carried through.
However, the presence of a mass gap in the system distinguishes between the
contributions of the macroscopic and microscopic excitations to the partition
function. While, for example, this observation does indeed put the $E_8$ based
character formulas and the structure of the integrable scaling field theory
of the Ising model in a magnetic field in the same framework, a better
understanding of this $q$-series away from criticality is lacking. We hope
to understand how the mass spectrum may be extracted by upgrading
the combinatoric arguments to the massive case, as well as some understanding
of how this character construction is related to Baxter's corner transfer
matrix method. It should also be noted that the thermodynamic
Bethe ansatz equations may be set up for models which do not necessarily
possess lattice Bethe ansatz integrability, and $q$-series modular forms
may thus be associated with any such model possessing a factorizable
$S$-matrix.

These combinatoric formulas have deep relations with several objects of
mathematical interest.
In \rComb,
bijective correspondences between the combinatorics of Young tableau and
the counting of Bethe ansatz states in $su(n)$ invariant systems have been
proposed. In particular,
they could identify an appropriate Lascoux-Sch\"utzenberger charge of tableaux
(see \rMac \rComb  for definitions)
in terms of the integer prescriptions to string solutions (what they called
rigged configurations) to get expressions for $q$-Kostka polynomials (\rMac).
The results presented here indicate that for restricted tableaux (with no
more than $l$ columns), this counting procedure relates the $q$-Kostka
polynomials of these infinite (for the filled fermi sea) tableaux with
modular forms associated with
affine Lie algebras. The connection with the combinatorics of
crystal bases \rcryone (an explicit construction has again been carried out
via a Fock space realization on extended Young tableaux \rcrytwo) should
provide insight into the relationship between the counting of solutions
to the Bethe
equations and the corner transfer matrix calculations mentioned above.
	There are, of course, several open questions we have mentioned,
and several directions of continuing research. It would be interesting
to see if the counting scheme is actually of some universal form and
whether the classification of states of other models may be brought into this
form. Two interesting cases are the $Z_n$ parafermions studied in \rAl
(through the ``zeroes of T'' of the Fateev-Zamolodchikov model \rFatZam
) and the model of \rBGS. The character formulas (at least for the $Z_n$
models can, of course, be obtained by level-rank duality \rKKMMtwo. A
study of the model in ref. \rBGS should yield interesting results.
The question being posed
is whether there is information beyond what is implicit in the formulas for
the branching functions that distinguishes between different models.
	The outstanding question is, of course, the validity of the
string hypohesis -- namely, do the solutions of \bethe actually fall into
these classes? If not, how does the counting procedure rearrange the
non-stringy effects to produce such clean results?

{\bf Acknowledgements}

I would like to thank Prof. H. Grosse and the organizers of the workshop
on 2-dimensional field theories at the Erwin Schr\"odinger Institute for
their hospitality and for providing an opportunity to present some early
version of these results in the first week of March.
I would also like to thank
C. Ahn, G. Albertini, A. Berkovich, V. K. Dobrev, A. Ganchev,
T. Jayaraman, R. Kedem, B.~M. McCoy, E. Melzer, W. Nahm, K. S. Narain,
V. Pasquier, F. Ravanini and A. Recknagel for useful
discussions and B. Razzaghe-Ashrafi and B.~M. McCoy for their comments on
an earlier version  of this manuscript.

\listrefs

\end